# The ferroelectric nematic phase: An optimum liquid crystal candidate for nonlinear optics

*C. L. Folcia, J. Ortega, R. Vidal, T. Sierra and J. Etxebarria**

Prof. C.L. Folcia [1], Dr. J. Ortega [1], R. Vidal [2], Dr. T. Sierra [2], Prof. J. Etxebarria [1]
[1] Department of Physics, Faculty of Science and Technology, UPV/EHU, 48080 Bilbao, Spain
[2] Instituto de Nanociencia y Materiales de Aragón (INMA), Química Orgánica, Facultad de Ciencias. CSIC-Universidad de Zaragoza, Pedro Cerbuna 12, ES-50009 Zaragoza, Spain
E-mail: (j.etxeba@ehu.eus)

Materials that exhibit high nonlinear optical (NLO) susceptibilities are considered as promising candidates for a wide range of photonic and electronic applications. Here we argue that the ferroelectric nematic ($N_F$) materials have sufficient potentialities to become materials for the next-generation of NLO devices. We have carried out a study of the efficiency of optical second-harmonic generation in a prototype $N_F$ material, finding a nonlinear susceptibility of 5.6 pmV$^{-1}$ in the transparent regime, one of the highest ever reported in ferroelectric liquid crystals. Given the fact that the studied molecule was not specifically designed for NLO applications we conclude there is still margin to obtain $N_F$ materials with enhanced properties that should allow their practical use.

## 1. Introduction

Nonlinear optical (NLO) effects can be used for a wide range of applications in electrooptics, photonics or biomedical technologies. Among them, second-order effects present remarkable interest and are currently used, e.g., for frequency doubling in lasers or in Pockels cells. In most applications available nowadays, the materials implemented in devices are classical inorganic crystals such as $LiNbO_3$, $KH_2PO_4$ (KDP) or $KTiOPO_4$ (KTP). However, organic materials have better potentialities for NLO applications since, compared to inorganic crystals, present more versatile synthesis and, therefore, it is easier to design *ad hoc* organic molecules for NLO purposes.[1,2] Among organic materials, liquid crystals (LC) are especially attractive since they present fluidity and are easily integrable in semiconductor devices. LCs would be a compelling alternative to inorganic materials if large NLO strengths were achieved.[3,4]

In order to obtain high second order optical effects in LCs, the molecules must incorporate long π-bridged donor-acceptor systems. However, macroscopic second-order NLO effects are only possible in non-centrosymmetric materials, which implies that, in practice, only ferroelectric (FLCs) liquid crystals can present second-order NLO properties. At molecular level, the optimum response in FLCs is achieved when long π-bridged donor-acceptor groups are set along the polarization direction. This requirement has been an important drawback in classical FLCs, since long π-bridged donor-acceptor systems incorporated along the molecular long axis are pointless, because the head to tail invariance prevents second-order responses. On the other hand, the incorporation of bulky chromophores along other directions can be incompatible with a large aspect ratio required for mesogeinity. Different approaches of molecular design have been proposed to obtain FLCs for NLO. The most outstanding designs are bent-core LCs[5-14] or azo bridged dimers and trimers.[15-18] In both cases relatively long chromophores can be integrated in the molecules with an acceptable component along the polarization direction **(Figure 1)**.

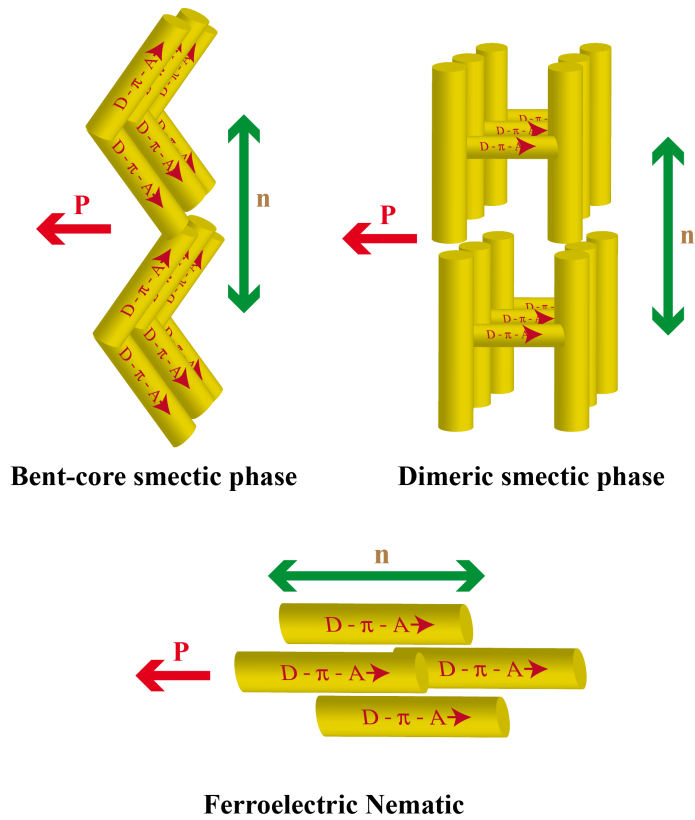

**Figure 1.** Scheme of the molecular arrangement in different FLC phases for NLO. **n** is the molecular director, **P** is the polarization and D-π-A denotes the donor-acceptor unit.

Recently the ferroelectric nematic ($N_F$) mesophase has been found out.[19-29] This phase is uniaxial and therefore spontaneous polarization appears along the director, which means that the head to tail invariance is no longer present. Therefore, contrary to previous FLCs, $N_F$ LCs offer the possibility of introducing long donor-acceptor groups along the long axis of the molecule (**Figure 1**), which turns out to be very advantageous. Moreover, the $N_F$ phase presents very high nematic order and it is reasonably easy to align in glass cells.[26,27,29] As mentioned by Chen et al,[27] in a $N_F$ phase, a uniform nearly saturated bulk polarization can be achieved just from surface alignment, even in the absence of an applied electric field. This is an important advantage with respect to other types of materials such as polymers, which require corona poling with strong electric fields to induce polarization. Another important advantage of $N_F$ LCs is the strong interaction with electric fields, which implies much more efficient responses than the usual quadrupolar coupling of the classical nematic (N) LCs. In fact, switching in $N_F$ LCs can be achieved by applying very low electric fields (~1Vcm$^{-1}$).[26,27,29] All of these properties make the $N_F$ phase an ideal candidate for NLO purposes.

In this work we investigate the efficiency for optical second harmonic generation (SHG) of the calamitic liquid crystal 4-[(4-nitrophenoxy)carbonyl]phenyl2,4-dimethoxybenzoate (RM734)[19,20] in the $N_F$ phase (**Figure 2a**). The study consists in the determination of the second order dielectric susceptibility $d_{ij}$ tensor from the characteristics of the emitted second harmonic light. The results have been interpreted at a microscopic level using available data of the molecular hyperpolarizability and the degree of polar order of the $N_F$ phase. Our motivation is to present a method to serve as a guide for the characterization of new $N_F$ materials with high SHG efficiencies. This seems of interest, especially if one keeps in mind that an efficient material for SHG is also efficient for other second order nonlinear effects such as the electronic Pockels effect, with obvious applications. Regarding RM734 and other $N_F$ materials, it is worth mentioning that a strong SHG light emission has already been reported,[30-32] although only incomplete quantitative data have been given in those works.

This manuscript is structured as follows. Firstly, the experimental data are presented. Next, the main SHG results are discussed, focusing on the connection between the macroscopic and microscopic properties of the material. Then, some conclusions are drawn and future prospects for the $N_F$ phase are commented. Finally, an experimental section is presented regarding the synthesis of the material and the description of the SHG setup.

## 2. Experimental results

The symmetry of the $N_F$ phase is $\infty m$, i.e., it is an optically uniaxial medium, and the second order dielectric susceptibility tensor under Kleinman condition is

$$d = \begin{pmatrix} 0 & 0 & 0 & 0 & d_{31} & 0 \\ 0 & 0 & 0 & d_{31} & 0 & 0 \\ d_{31} & d_{31} & d_{33} & 0 & 0 & 0 \end{pmatrix}, \tag{1}$$

where the $Z$ axis is along the polar direction. Here the three-index notation of the susceptibility has been contracted to two indices in the usual way. There are only two independent coefficients, $d_{33}$ and $d_{31}$. SHG light generated along $Z$ (extraordinary polarization) is driven by the $d_{33}$ coefficient for extraordinary fundamental light, and by $d_{31}$ for ordinary fundamental light. Both coefficients can then be determined independently by selecting the polarization of the incident light as extraordinary and ordinary, and measuring in both cases the extraordinary SHG light **(Figure 2b)**.

The SHG power for fundamental light at normal incidence with power $P^\omega$ is given by the expression

$$P^{2\omega} = C(t^{2\omega})^2(t^\omega)^4(P^\omega)^2 d_{eff}^2 \frac{\sin^2\left(\frac{2\pi}{\lambda}\Delta n_d L\right)}{\left(\frac{2\pi}{\lambda}\Delta n_d\right)^2}, \tag{2}$$

where $t^\omega$ and $t^{2\omega}$ are standard transmission Fresnel coefficients, $\lambda$ is the wavelength of the incident light, $L$ is the sample thickness, $\Delta n_d = n^{2\omega} - n^\omega$ is the dispersion of the material, and $d_{eff}$ is $d_{33}$ or $d_{31}$ in our experiment. The constant $C$ depends on the experimental device and is determined from the SHG intensity of a known standard sample (see **Experimental section**). It is evident from the expression of $P^{2\omega}$ that the determination of $d_{eff}$ inevitably requires the knowledge of the dispersion $\Delta n_d$.

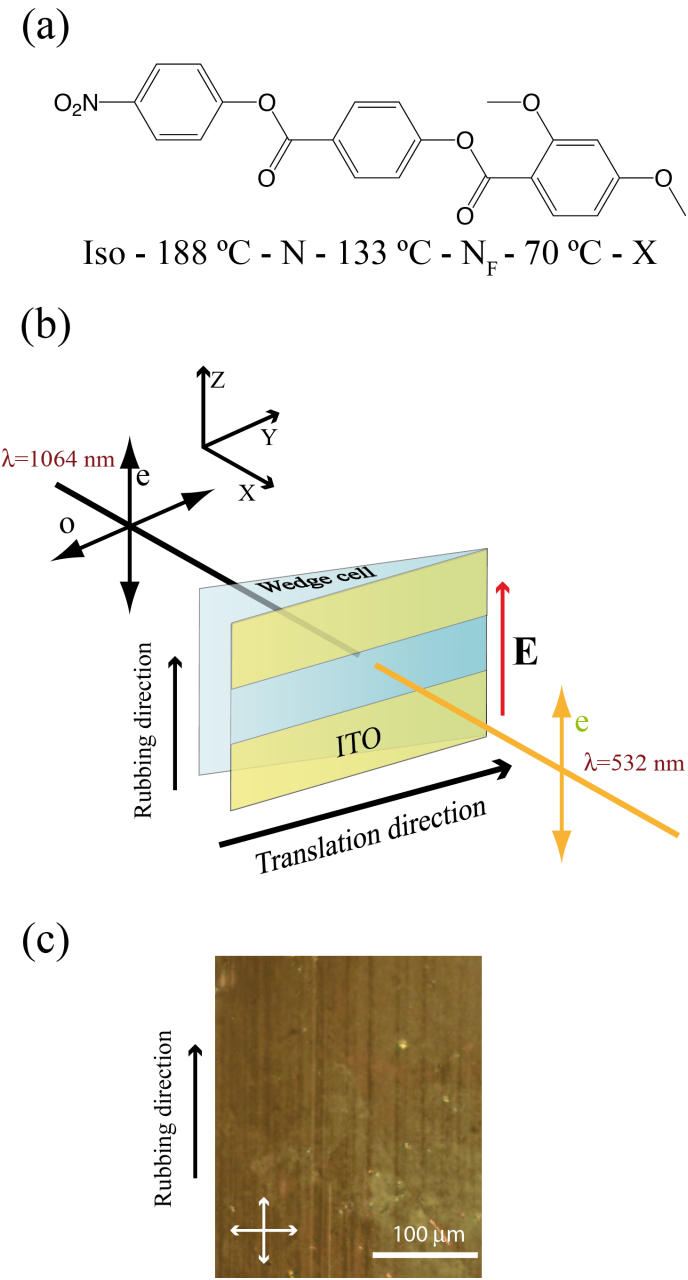

**Figure 2.** (a) Structure and phase sequence on cooling of RM734. (b) Scheme of the wedge cell and the different configurations for the SHG measurements. (c) Texture of the $N_F$ phase under crossed polarizers at a temperature of 118ºC. White arrows indicate polarizer directions.

A wedge sample inside a high temperature stage was situated on the displacement station with the extraordinary axis, in the vertical direction (see **Experimental section**). Thus, the incident light was vertically (horizontally) polarized for the measurement of $d_{33}$ ($d_{31}$), and the SHG light was vertically polarized in both cases. The experiment consisted in measuring $P^{2\omega}$ at normal incidence while the sample was displaced perpendicularly to the incident light so that the sample thickness $L$ was continuously varied (**Figure 2b**). In this way we obtain a sinusoidal curve of $P^{2\omega}$ as a function of the transverse displacement (Maker fringes). The $sin^2\left(\frac{2\pi}{\lambda}\Delta n_d L\right)$ dependence of $P^{2\omega}$ allows to determine $\Delta n_d$ from the periodicity, and $d_{33}$ or $d_{31}$ are obtained from the maxima of the curve.

The cell was capillarity filled at about 15 ºC below the clearing point in the N phase. In contrast to previous reports,[26,27,29] we were surprised by the fact that a dark texture typical of a homeotropic alignment was observed in our sample. The same texture appeared with non-wedged cells of 12 and 20 μm. A drastic texture change occurred at about 133 ºC when the transition to the $N_F$ phase takes place: A non-defined texture with regions that extinguish in different directions was observed. This texture becomes nearly homogeneous, with extinctions parallel and perpendicular to the rubbing direction, when a field of 0.5 to 1 Vcm$^{-1}$ is applied, irrespective of the field polarity, and a somewhat less defined texture but with the same extinction directions appears after removing the field (**Figure 2c**).

Before going into the $d_{ij}$ results, we analyze qualitatively some general aspects of the SHG light. An intense signal linearly polarized in the extraordinary (vertical) direction was observed for extraordinary incident light. In this case the SHG field must be proportional to $d_{33}E_e^\omega E_e^\omega$, where the index *e* means extraordinary. No difference was noted for opposite field polarities. The signal without field was also similar although in general somewhat weaker depending on the illuminated region of the sample, which means that the applied field favors a better alignment of the material. The SHG signal under field saturated for fields of about 2

Vcm⁻¹. Our study was performed under a field of 2.6 Vcm⁻¹. On the other hand, a much weaker signal with a rather small signal-to-noise ratio was detected when the incident light was polarized in the ordinary (horizontal) direction. In this case the SHG field is proportional to $d_{31}E_o^\omega E_o^\omega$, where $o$ means ordinary. Accordingly, as expected from the molecular geometry, $d_{33} \gg d_{31}$. We will return to this point later.

**Figure 3a** shows the SHG intensity as a function of the thickness L for extraordinary fundamental light at 118 ºC. The curve was fitted to a $Asin^2\left(\frac{2\pi}{\lambda}\Delta n_d^{ee}L\right)$ dependence from which $\Delta n_d^{ee} = n_e^{2\omega} - n_e^\omega = 0.064$ results. The coefficient $d_{33}$ was determined from the amplitude $A$ and resulted to be $d_{33} = 5.6$ pmV⁻¹ at this temperature. **Figure 3b** shows the temperature dependence of the SHG intensity in the N$_F$ phase. The measurement was carried out on heating, situating the sample on one of the maxima like those in **Figure 3a**. The coefficient $d_{33}$ is proportional to the square root of the SHG intensity under the reasonable assumption that the temperature dependence of $\Delta n_d^{ee}$ may be neglected.

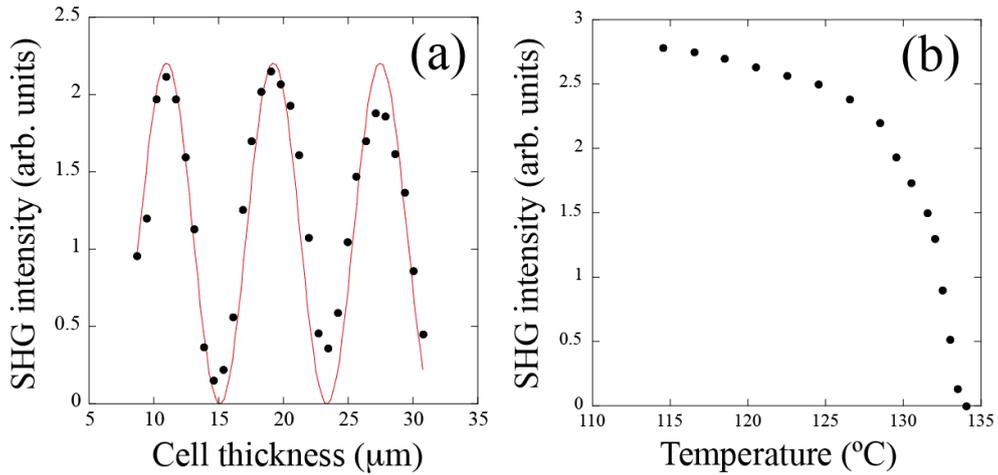

**Figure 3.** (a) SHG intensity vs. wedge cell thickness. Red line is a fit to $Asin^2\left(\frac{2\pi}{\lambda}\Delta n_d^{ee}L\right)$. (b) SHG vs. temperature. Data were collected on heating up to the N phase.

As mentioned above, the coefficient $d_{31}$ can be measured illuminating the sample with ordinary (horizontally polarized) light. However, the accurate determination of this coefficient was difficult because of the high $d_{33}/d_{31}$ ratio. As a consequence of this, unavoidable misalignments of the sample cause the intrusion of the SHG signal coming from $d_{33}$. In view of this, we selected a sample area with optimal alignment for the measurements. The detected signal was about 100 times weaker than the one observed at a maximum of Figure 3a, meaning that a maximum limit of about $d_{33}/10$ (0.6 pmV$^{-1}$) can be established for $d_{31}$.

## 3. Discussion of the results

The value of the $d_{33}$ coefficient (5.6 pmV$^{-1}$) of RM734 is similar to those of the most efficient LC reported up to know. This fact is quite remarkable since RM734 LC has not been designed for NLO purposes and, therefore, does not include efficient donor-acceptor groups. In fact, the most efficient LC reported up to now is only slightly higher and it presents a maximum value for the $d_{ij}$ coefficients of 6.8 pmV$^{-1}$ in transparent regime.[18] A compilation of the most outstanding LCs for NLO applications reported so far can be found in the table 3 of the same reference.

We now try to connect the macroscopic NLO coefficients $d_{ij}$ with the molecular and structural parameters. We use the oriented-gas model and assume that the hyperpolarizability of the molecule comes essentially from the NO$_2$-π-O molecular segment, since the rest of the molecule does not possess any group with relevant donor or acceptor character. Besides, the ester groups connecting the benzene rings interrupt the possibility of charge transfer along the molecular long axis, restricting the donor-acceptor conjugation length to only one ring (the leftmost ring in **Figure 1a**). Thus, it seems reasonable to assume that the hyperpolarizabilty has simply one component directed along the molecular long axis $z$ ($\beta = \beta_{zzz}$), with a magnitude similar to that of the fragment that contains the donor-π-acceptor group, i.e., the *p*-nitrophenol molecule.

For *p*-nitrophenol, experimental data using the EFISH technique are available. At a fundamental wavelength of 1.907 μm, reference [33] gives $\beta = 3.0 \times 10^{-30}\ esu$ along the donor-acceptor direction. For our excitation wavelength $\lambda = 1.064$ μm this value is slightly resonance enhanced. Using the two-level dispersion model we obtain $\beta = 4.2 \times 10^{-30}$ esu, taking a resonance wavelength $\lambda_{max} = 0.304$ μm.[33]

We will write the susceptibility tensor in a laboratory reference system **(Figure 4a)** where the *Z* axis is along the macroscopic polar axis of the $N_F$ phase. Within the oriented-gas model, the $d_{IJK}$ components of the susceptibility tensor in this frame are given by

$$d_{IJK} = Nf^3 \langle \beta_{IJK} \rangle, \qquad (3)$$

where $\langle \beta_{IJK} \rangle$ are the thermally averaged hyperpolarizability tensor components in the laboratory system, *N* is the number of molecules per unit volume, and *f* the Lorentz factor (we assume the same factors for the second-harmonic and fundamental radiation frequencies, and take $f = (n^2 + 2)/3$ with an average refractive index *n* =1.6). *N* can be obtained from the mass density $\rho = 1.3\ \text{gcm}^{-3}$ (see reference [26]) as $N = N_A \rho/M = 1.85 \times 10^{21}\ \text{molecules.cm}^{-3}$, where *M* is the molar mass and $N_A$ the Avogadro number.

The $\langle \beta_{IJK} \rangle$ components are easily obtained by a simple transformation of axes and subsequent thermal averaging. A straightforward calculation gives rise to a final tensor which has the shape given by Eq. (1) in the two-index Voigt notation. The two independent components are

$$d_{33} = Nf^3\langle\cos^3\theta\rangle\beta$$

$$d_{31} = Nf^3\langle\cos\theta\sin^2\theta\rangle\beta/2$$

(4)

where it has been taken into account that the azimuthal angle $\phi$ distribution of the molecular long axes is isotropic. On the other hand, the distribution function for the polar angle $\theta$ is $F(\theta) = A\exp[-U(\theta)/k_BT]$ where $A$ is a normalization constant, $k_B$ the Boltzmann constant, and $U(\theta)$ the nematic potential, which in its simplest form can be expressed in terms of two functions of the temperature $T$, $a(T)$ and $b(T)$, as

$$U(\theta) = a(T)\sin^2\theta + b(T)\sin^2(\theta/2). \tag{5}$$

Here $a(T)$ accounts for the intensity of the usual nematic potential (which is taken as proportional to the nematic order parameter in the Maier-Saupe theory), whereas $b(T)$ gives the strength of a new contribution to the nematic potential due to the existence of ferroelectricity.

Now we will obtain the numerical values of $a(T)/k_BT$ and $b(T)/k_BT$ at 118 °C (where the SHG mesurements were performed) from two experimental data:

i) The polarization value at that temperature, $P(118) = 5.1\ \mu\text{Ccm}^{-2}$.[26] This gives $\langle\cos\theta\rangle = P(118)/N\mu = 0.723$, where $\mu = 11.4$ D is the molecular dipole moment.[26]

ii) The ratio between the actual birefringence at 118 °C ($\Delta n$) and the extrapolated birefringence at the same temperature in the absence of polar order ($\Delta n_{NP}$), $\Delta n/\Delta n_{NP} = 0.244/0.228$ (see Figure S17 in [26]). Since the birefringence is expected to be proportional to the nematic order parameter $S = (3\langle\cos^2\theta\rangle - 1)/2$, this birefringence ratio can be connected with $S/S_{NP}$, where $S_{NP}$ is the order parameter deduced from the distribution function with $b(T) = 0$.

These two conditions for $\langle \cos\theta \rangle$ and $\langle \cos^2\theta \rangle$ allow to deduce numerically $a(T)/k_B T = 4.97$ and $b(T)/k_B T = 2.62$, at 118 °C. The resulting $U(\theta)/k_B T$ function and the distribution function $F(\theta)$ are plotted in **Figure 4b**.

Finally, to obtain the $d_{ij}$ values we calculate the thermal averages $\langle \cos^3\theta \rangle = 0.605$, and $\langle \cos\theta \sin^2\theta \rangle = 0.118$, which give $d_{33} = 6.9$ pmV$^{-1}$ and $d_{33} = 0.66$ pmV$^{-1}$. The nematic order parameter results $S = 0.694$. Given the simplicity of the model and the number of approximations involved, the agreement with the experimental results $d_{33} = 5.6$ pmV$^{-1}$ and $d_{31} < 0.6$ pmV$^{-1}$ is certainly remarkable.

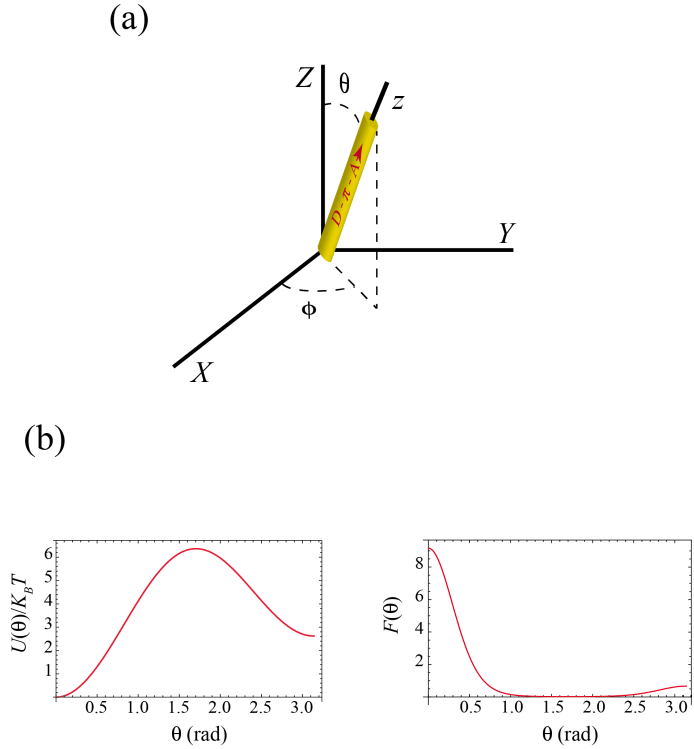

**Figure 4.** (a) Scheme of the molecular and laboratory reference frames. (b) Nematic potential $U(\theta)/k_B T$ and normalized angular distribution function $F(\theta) = A\exp[-U(\theta)/k_B T]$ at 118 °C in the N$_F$ phase.

## 4. Conclusions

SHG measurements of RM734 give a $d_{33}$ coefficient of 5.6 pmV$^{-1}$ in the N$_F$ phase, one of the largest NLO coefficients reported to date for FLCs. This value is in the verge to enable viable applications in NLO and electrooptics. We have also seen that the size of the NLO susceptibility is proportional to the molecular hyperpolarizability $\beta$ along the donor-acceptor direction. This means that it is still possible to substantially increase the susceptibility values if molecules are used where the charge-transfer channel along the long molecular axis is not interrupted, as is the case in the current molecule. As an example, a N$_F$ liquid crystal made of molecules with a $\beta$ value equal to that of the standard chromophore DR-1, would present a $d_{33}$ coefficient higher than 100 pmV$^{-1}$. In this sense, it is interesting to note that $\beta$ has a quadratic or greater than quadratic dependence with the distance between the donor and the acceptor. Obviously, the molecular design that allows for the electronic conjugation should be compatible with the possibility of keeping the N$_F$ phase. Another attractive possibility seems to dissolve NLO chromophores in N$_F$ materials.[34] If the admitted chromophore concentration is high enough without causing N$_F$ phase disruption, we believe that these structures can lead to novel applications that may revolutionize the technology based on nematic materials.

## 5. Experimental section

**5.1. Synthesis**

RM734 was synthetized following previously reported procedures[20,26]. Experimental details concerning its preparation and characterization are included in the supporting information.

## 5.2. SHG measurements

The SHG experimental set up was described in detail elsewhere.[35] Briefly, polarized light from a Q-switched pulsed Nd:YAG laser ($\lambda = 1.064$ μm) illuminates the sample situated on a high resolution displacement station. The beam was collimated and its diameter was limited by a pinhole to be less than 1 mm diameter. The intensity of the emitted second harmonic light ($\lambda = 0.532$ μm) was measured by a photomultiplier after passing an analyzer. A wedge cell was prepared with both glasses treated with polyvinyl alcohol and parallel rubbed (*SYNPOLAR* cell according to the definition of reference [27]), defining the extraordinary axis of the optical medium. ITO strip electrodes spaced 5 mm apart allow for the application of an in-plane electric DC field parallel or antiparallel to the rubbing direction. The thickness of the cell at different points was measured using an interferometric method allowing for a precise knowledge of the wedge angle (3.1 x 10$^{-3}$ rad).

The calibration of the setup was carried out using a y-cut quartz plate of 0.7 mm of thickness. The plate was rotated about the (vertical) *x*-axis while the SHG was measured for *x*-polarized *ω* and 2*ω* lights (ordinary-to-ordinary conversion). The SHG intensity changes as a function of the angle of incidence showing maxima and minima. From the size of the first maximum and using the quartz data ($d_{11} = 0.4$ pmV$^{-1}$ and $\Delta n_d = n_o^{2\omega} - n_o^{\omega} = 0.013$) the constant *C* in Eq. (2) is easily extracted.

**Acknowledgements**

This work was supported by the Euskal Herriko Unibertsitatea [Project GIU18/146], the MICIU/AEI/FEDER, EU funds [project PGC2018-093761-B-C31] and the Gobierno de Aragón-FSE [E47_20R]. The authors would like to thank the "Servicios Científico Técnicos" of CEQMA, CSIC-Universidad de Zaragoza.

References:

[1] P. N. Prasad, D. J. Williams, *Introduction to Nonlinear Optical Effects in Molecules and Polymers*, Wiley, New York (1991), pp. 132-174.

[2] Ch. Bosshard, K. Sutter, Ph. Pretre, J. Hulliger M. Florsheimer, P. Kaatz and P. Günter, *Organic nonlinear optical materials*, Gordon and Breach publishers, Switzerland, 1995.

[3] D. M. Walba, D. J. Dyer, P. L. Cobben, T. Sierra, J. A. Rego, C. A. Liberko, R. Shao, N. A. Clark, *Mat. Res. Soc. Symp. Proc.* **1995**, 392, 157.

[4] Y. Zhang, J. Etxebarria, *Ferroelectric liquid crystals for nonlinear optical applications*. In *Liquid crystals beyond displays*; Li, Quan, Ed.; Wiley: Hoboken, NJ, **2012**, 111–156.

[5] S. W. Choi, Y. Kinoshita, B. Park, H. Takezoe, T. Niori, *Jpn. J. Appl. Phys.* **1998**, 37, 3408.

[6] F. Kentischer, R. Macdonald, P. Warnick, G. Heppke, *Liq. Cryst.* **1998**, 25, 34.

[7] J. Ortega, N. Pereda, C. L. Folcia, J. Etxebarria, M. B. Ros, *Phys. Rev. E* **63**, **2000**, 011702.

[8] J. A. Gallastegui, C. L. Folcia, J. Etxebarria, J. Ortega, I. De Francisco, M. B. Ros, *Liq. Cryst.* **2002**, 29, 1329.

[9] F. Araoka, H. Hoshi, H. Takezoe, *Phys. Rev. E* **2004**, 69, 051704.

[10] J. Ortega, J. A. Gallastegui, C. L. Folcia, J. Etxebarria, N. Gimeno, M. B. Ros, *Liq. Cryst.* **2004**, 31, 579.

[11] H. Takezoe, Y. Takanishi, *Jpn. J. Appl. Phys.* **2006**, 45, 597.

[12] J. Martínez-Perdiguero, I. Alonso, C. L. Folcia, J. Etxebarria, J. Ortega, *J. Mater. Chem.* **2009**, 19, 5161.

[13] I. C. Pintre, J. L. Serrano, M. B. Ros, J. Martínez-Perdiguero, I. Alonso, J. Ortega, C. L. Folcia, J. Etxebarria, R. Alicante, B. Villacampa, *J. Mater. Chem.* **2010**, 20, 2965.

[14] B. Champagne, J. Guthmuller, F. Perreault, A. Soldera, *J. Phys. Chem. C* **2012**, 116, 7552.

[15] D. M. Walba, D. J. Dyer, T. Sierra, P. L. Cobben, R. Shao, N. A. Clark, *J. Am. Chem. Soc.* **1996**, 118, 1211.

[16] D. M. Walba, L. Xiao, P. Keller, R. Shao, D. Link, N. A. Clark, *Pure Appl. Chem.* **1999**, 71, 2117.

[17] Y. Zhang, J. Martinez-Perdiguero, U. Baumeister, C. Walker, J. Etxebarria, M. Prehm, J. Ortega, C. Tschierske, M. J. O'Callaghan, A. Harant, M. Handschy, *J. Am. Chem. Soc.* **2009**, 131, 18386.

[18] Y. Zhang, J. Ortega, U. Baumeister, C. L. Folcia, G. Sanz-Enguita, C. Walker, S. Rodríguez-Conde, J. Etxebarria, M. J. O'Callaghan, K. More, *J. Am. Chem. Soc.* **2012**, 134, 16298.

[19] R. J. Mandle, S. J. Cowling, J. W. Goodby, *Phys.Chem.Chem.Phys.* **2017**, 19, 11429.

[20] R. J. Mandle, S. J. Cowling, J. W. Goodby, *Chem. Eur. J.* **2017**, 23, 14554.

[21] H. Nishikawa, K. Shiroshita, H. Higuchi, Y. Okumura, Y. Haseba, S. Yamamoto, K. Sago, H. Kikuchi, *Adv. Mater.* **2017**, 29, 1702354.

[22] A. Mertelj, L. Cmok, N. Sebastián, R. J. Mandle, R. R. Parker, A. C. Whitwood, J. W. Goodby, M. Čopič, *Phys. Rev. X* **2018**, 8, 41025.

[23] R. J. Mandle, A. Mertelj, *Phys. Chem. Chem. Phys.* **2019**, 21, 18769.

[24] N. Sebastián, L. Cmok, R. J. Mandle, M. R. de la Fuente, I. Drevenšek-Olenik, M. Čopič, A. Mertelj, *Phys. Rev. Lett.* **2020**, 124, 37801.

[25] P. L. M. Connor, R. J. Mandle, *Soft Matter* **2020**, 16, 324.

[26] X. Chen, E. Korblova, D. Dong, X. Wei, R. Shao, L. Radzihovsky, M. A. Glaser, J. E. Maclennan, D. Bedrov, D. M. Walba, N. A. Clark, *PNAS* **2020**, 117, 14021.

[27] X. Chen, E. Korblova, M. A. Glaser, J. E. Maclennan, D. M. Walba, N. A. Clark, *PNAS* **2021**, 118, e2104092118.

[28] O. D. Lavrentovich, *PNAS* **2020**, 117, 14629.

[29] F. Caimi, G. Nava, R. Barboza, N. A. Clark, E. Korblova, D. M. Walba, T. Bellini, L. Lucchetti, *Soft Matter* **2021**, 17, 8130.

[30] J. Li, H. Nishikawa, J. Kougo, J. Zhou, S. Dai, W. Tang, X. Zhao, Y. Hisai, M. Huang, S. Aya, *Sci. Adv.* **2021**, 7, eabf5047.

[31] N. Sebastián, R. J. Mandle, A. Petelina, A. Eremin, A. Mertelj, *Liq. Cryst. (in press)*, DOI: 10.1080/02678292.2021.1955417.

[32] X. Zhao, J. Zhou, H. Nishikawa, J. Li, J. Kougo, Z. Wan, M. Huang, S. Aya. *PNAS.* **2021**, 118, e2111101118.

[33] L. T. Cheng, W. Tam, S. H. Stevenson, G. R. Meredith, G. Rikken, S. R. Marder, *J. Phys. Chem.* **1991**, 95, 10631.

[34] X. Chen, Z. Zhu, M. J. Magrini, E. Korblova, C. S. Park, M. A. Glaser, J. E. Maclennan, D. M. Walba, N. A. Clark, arXiv:2110.10826

[35] N. Pereda, C. L. Folcia, J. Etxebarria, J. Ortega, M. B. Ros, *Liq. Cryst.* **1998**, 24, 451.

# The ferroelectric nematic phase: An optimum liquid crystal candidate for nonlinear optics

*C. L. Folcia, J. Ortega, R. Vidal, T. Sierra and J. Etxebarria\**

The synthetic route to RM734 was adapted from the procedures described in (references 20 and 26 in the manuscript), and is depicted in Scheme 1.

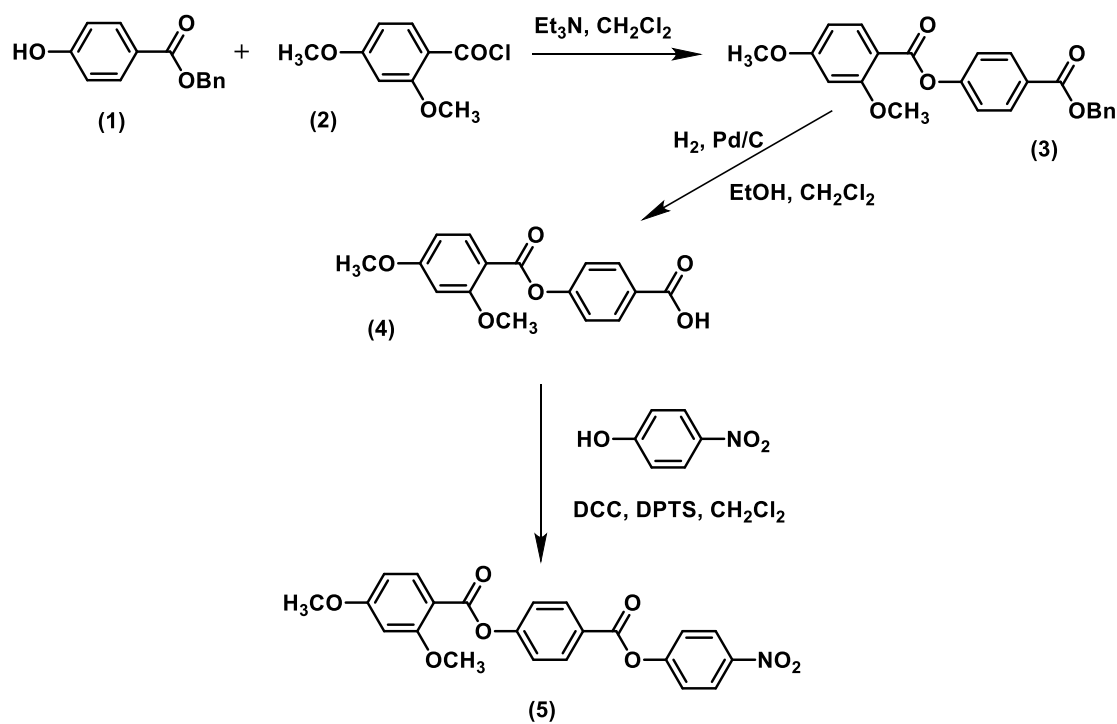

Scheme 1. Synthetic route followed to prepare RM734 (**5**).

*4-[(benzyloxy)carbonyl]phenyl 2,4-dimethoxybenzoate (3)*

A solution of benzyl 4-hydroxybenzoate (**1**) (2.8 g, 12.5 mmol) and 2,4-dimethoxybenzoyl chloride (**2**) (2.5 g, 12.5 mmol) in 100 mL of dry CH2Cl2 (100 mL) was cooled to 0°C, and triethylamine (1.5 g, 15 mmol, 2.1 mL) was then added dropwise. After stirring overnight at r.t., the reaction mixture was poured into a saturated aqueous solution of NH4Cl (100 mL) and extracted with CH2Cl2 (3x50 mL). The combined organic layers were washed with water (3x60 mL), brine (3x60mL), dried over MgSO4, filtered and concentrated at reduced pressure. The crude product was purified by flash chromatography, using silica gel and CH2Cl2 as eluent. Compound **3** was thus obtained as a white solid (2.6 g, 54%).

*4-(2,4-dimethoxybenzoyloxy)benzoic acid (4)*

A solution of the benzyl-protected carboxylic acid (**3**) (2.6 g, 6.7 mmol) in CH2Cl2 (20 mL) and EtOH (20 mL) was evacuated and purged with argon. Then, 20% Pd/C catalyst (0.52 g) was added and the argon atmosphere was replaced by hydrogen gas after three vacuum-hydrogen cycles. After two-hours stirring at r.t., hydrogen was evacuated from of the reaction flask. The reaction mixture was filtered through a Celite pad, which was further washed with butanone. After evaporation of the solvents under reduced pressure, acid **4** was obtained as a white powder (1.9 g 96%), which was used in the next step without further purification.

*4-[(4-nitrophenoxy)carbonyl]phenyl 2,4-dimethoxybenzoate (5)*

A suspension of compound **4** (1.40 g, 4.6 mmol) and 4-nitrophenol (**5**) (0.71 g, 5.1 mmol) in CH2Cl2 (50 mL) was prepared, and DCC (2.10 g, 10.2 mmol) and DPTS (0.68 g, 2.3 mmol) were added. After stirring for three days at r.t., the reaction mixture was filtered, and the resulting solution was washed with water, 5% acetic acid, water (3x60mL), brine (3x60mL), and then dried over MgSO4, filtered, and concentrated at reduced pressure. The resulting product was purified by flash chromatography (silica gel, CH2Cl2/2%EtOAc) followed by two subsequent crystallizations from EtOH and CH3CN. The pure product **5** (RM734) was obtained as a white crystalline solid (0.88 g 41%).

$^1$H NMR (400 MHz, CDCl3) ∂ ppm: 8.36 - 8.32 (m, 2H; Ar*H*), 8.27 - 8.24 (m, 2H; Ar*H*), 8.10 (d, *J* = 8.7 Hz, 1H; Ar*H*) 7.45 -7.41 (m, 2H; Ar*H*), 7.41 -7.37 (m, 2H; Ar*H*), 6.58 (dd, *J* = 2.3 Hz, *J* = 8.7 Hz, 1H; Ar*H*), 6.55 (d, *J* = 2.3 Hz, 1H; Ar*H*), 3.94 (s, 3H; OC*H$_3$*), 3.91 (s, 3H; OC*H$_3$*). $^{13}$C NMR (101 MHz, CDCl3) ∂ ppm: 165.53, 163.77, 162.91, 162.68, 156.13, 155.84, 145.55, 134.78, 131.98, 125.71, 125.43, 122.80, 122.65, 110.42, 105.13, 99.16, 56.20, 55.78. MS (ESI +): 446.0845 [M+Na]$^+$. Elemental analysis calcd. for C$_{22}$H$_{17}$NO$_8$: C 62.41, H 4.05, N 3.31, found: C 62.15, H 4.18, N 3.43.

## [1]H NMR

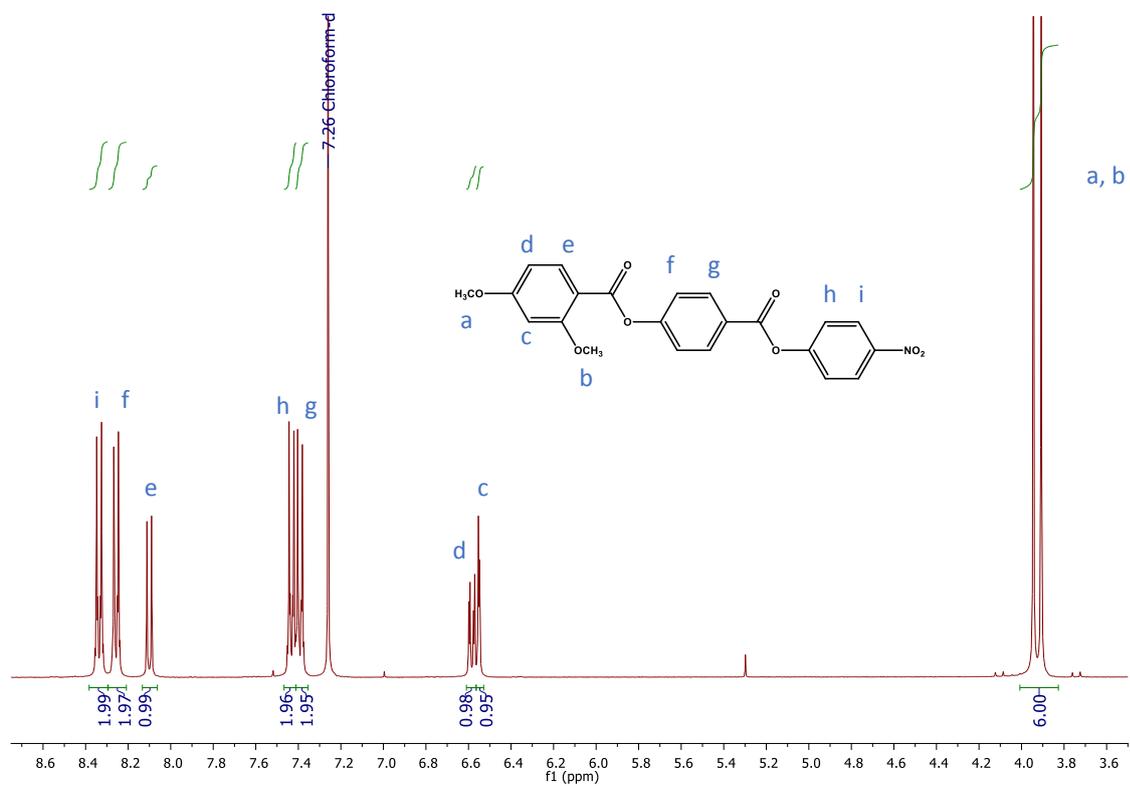

[1]H RMN spectrum of *4-[(4-nitrophenoxy)carbonyl]phenyl 2,4-dimethoxybenzoate* **(5)** in

CDCl$_3$

*<ins>13C NMR</ins>*

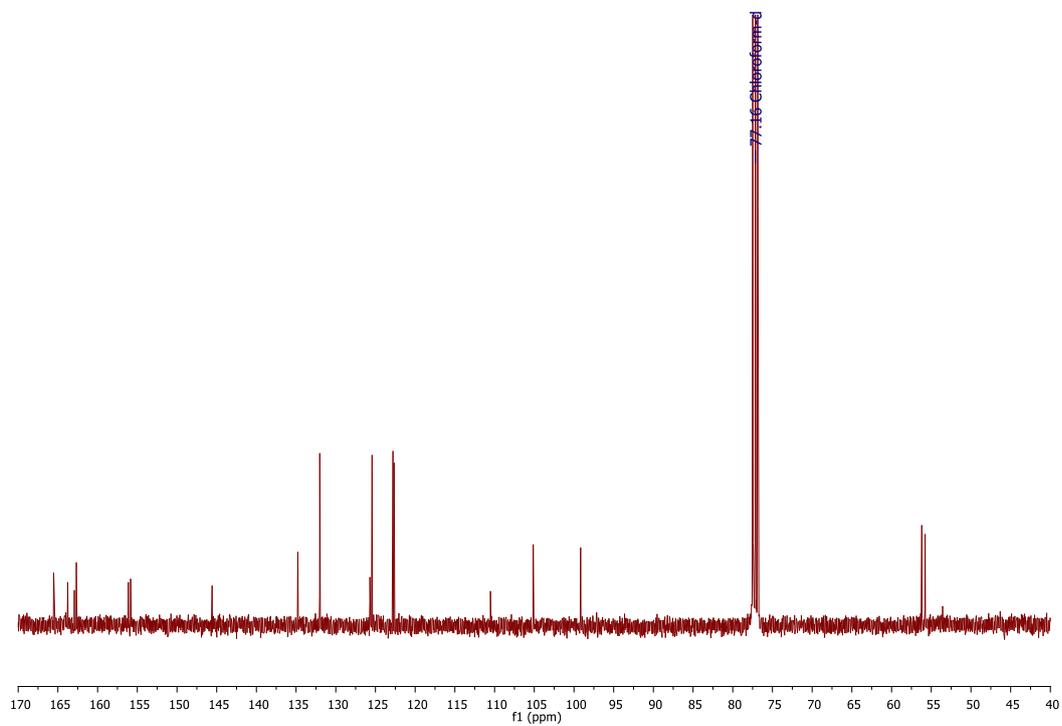

13C RMN spectrum of *4-[(4-nitrophenoxy)carbonyl]phenyl 2,4-dimethoxybenzoate (5)* in CDCl$_3$